\documentclass[preprint,12pt]{aastex}
\usepackage{emulateapj5}

\newcommand{\kms}{\mbox{km s$^{-1}$}}

\newcommand{\tk}{\mbox{$T_K$}}

\newcommand{\mean}[1]{\mbox{$\langle#1\rangle$}} 

\input{epsf}

\shorttitle{Inflow in Massive Regions}
\shortauthors{Wu \& Evans }
\begin{document}
\title {\bf Indications of Inflow Motions in Regions Forming Massive Stars}
\author {Jingwen Wu and Neal J. Evans II}
\affil{Department of Astronomy, The University of Texas at Austin,
        1 University Station, C1400, Austin, Texas 78712--0259}
\email{jingwen@astro.as.utexas.edu}
\email{nje@astro.as.utexas.edu}

\begin{abstract}
Observational evidence for inflowing motions in massive star forming
regions has been extremely rare.
We have made a spectroscopic survey of a sample of 28 massive star forming
cores associated with water masers. An
optically thick line of HCN 3$-$2 was used in combination with
optically thin lines (H$^{13}$CN 3$-$2 or C$^{34}$S 5$-$4, 3$-$2, and 2$-$1 ),
to identify ``blue'' line profiles that can indicate inflow.
Comparing intensities for 18 double-peaked line profiles
yields 11 blue and 3 red profiles that are statistically significant.
In the full sample of 28 sources, 
12 show blue profiles and 6 show red profiles that
are statistically significant based on the velocity offsets of lines
that are optically thick from those that are optically thin.
These results indicate that HCN $3-2$ emission may trace inflow in regions
forming high-mass stars.

\end{abstract}
\section{Introduction}
Gravitational collapse is a widely accepted theoretical
explanation of star formation, but the observational evidence for collapse has
long been controversial (e.g., Evans 1991, Myers et al. 2000). The major
difficulty is that collapse proceeds at a relatively low velocity
and is easily masked by other motions (e.g., outflows,
rotation and turbulence) in the cloud cores.
A general prediction of collapse models is a ``blue profile,''
a line asymmetry with the peak skewed to the blue side
in optically thick lines, while an optically thin line must peak at the
absorption part (usually a dip) of the optically thick line to rule out
the possibility of two velocity components.

For an individual source, one must exclude rotation or
outflow blobs as the source of the blue profile by mapping the source.
However, the predominance of blue profiles in a survey,
based on some objective criteria, can be an indication that inflow is
a statistically likely explanation. We use the term ``inflow" to
distinguish evidence of inflowing motions from a claim of gravitational
collapse or infall.
A profile that survives these tests provides a strong indication
of inflow, and the source can be seen as a collapse candidate.
To become a credible example of gravitational collapse,
gravity must be a plausible source of the motions, and the line
profiles should agree with some detailed model (Evans 2003).

Progress has been made in the last decade in studying low mass star forming
regions, triggered by
the observation of blue profiles toward B335 in lines of H$_{2}$CO and CS
and later the fitting (Zhou 1993, 1994, Choi et al. 1995)
of these lines with an inside-out collapse model (Shu 1977).
Systematic surveys have been made and predominantly blue profiles have been
found toward Class 0 (Gregersen et al. 1997, Mardones et al. 1997),
Class I (Gregersen et al. 2000),
and Class $-1$ (also known as pre-protostellar cores) sources 
(Gregersen $\&$ Evans 2000, Lee, Myers, \& Tafalla 1999, 2001).
Evidence of collapse has been increasingly accepted
in low mass star forming regions.

Evidence of inflow has been elusive in massive star forming regions,
partly because of the extremely complex and turbulent environment in
massive clouds.
Spectral inflow signatures have been
reported in a few sources including NGC 2264 IRS
(Wolf-Chase \& Gregersen 1997), W49 (Welch et al. 1988, Dickel \& Auer 1994),
G10.6-0.4 (Keto et al. 1988) and W51 (Zhang \& Ho 1997, Zhang et al. 1998).
However, surveys have not revealed statistical evidence of inflow
(e.g., Williams \& Myers 1999).

The characteristic blue profile will appear only if the
molecular tracer has a suitable optical depth and critical density. For
example, using moderately opaque tracers H$_{2}$CO and CS, which are good
tracers of blue profiles in Class 0 sources, Mardones et al. (1997)
found a low fraction of blue profiles in a sample of Class I sources.
With a more opaque line of HCO$^{+}$ 3$-$2, Gregersen et al. (2000)
detected a blue fraction in Class I sources
comparable to that in Class 0 sources.
The inflow tracers used in low mass cores may no longer be suitable
to probe inflow motion in high mass cores, which are typically 100 times
denser at a given radius than low mass cores (Mueller et al. 2002).
In a recent survey for blue profiles in high mass star forming regions,
Williams $\&$ Myers (1999) used CS 2$-$1
(critical density $n_{c}$ is 3.9$\times$ 10$^{5}$ cm$^{-3}$ at $\tk = 100$ K)
as the optically thick line.
Three out of 19 sources showed self-absorption features,
1 of which had a blue profile. We have used the HCN 3$-$2 line, which
may be more opaque in denser regions
($n_{c}=6.8\times10^{7}$ cm$^{-3}$ at $\tk =100$ K) than CS $2-1$.

\section{Observations}

The sample is a subset of a larger sample of dense cores
associated with H$_{2}$O masers that have been surveyed with several CS
transitions (Plume et al. 1992, 1997) and mapped in CS and dust emission
(Shirley et al. 2003, Mueller et al. 2002). The sources mapped in CS 5-4 
(Shirley et al. 2003) have virial masses within the nominal core radius 
(R$_{CS}$) ranging from 30 M$_{\odot}$ to 2750 M$_{\odot}$, 
with a mean of 920 M$_{\odot}$. That sample was selected to have
CS 7$-$6 emission with $T^{*}_{R} \geq 1$ K. The sample 
for HCN observations was selected randomly from the sample in Shirley et al. 
(2003); we were trying to cover all the range of mass in that sample. 
This sample has a mean virial mass of 800 M$_{\odot}$.
We used HCN 3$-$2 lines as the optically thick lines. H$^{13}$CN 3$-$2, or 
C$^{34}$S 5$-$4, 3$-$2, and 2$-$1 lines were used as optically
thin lines. For comparision, CS 5$-$4 lines were also taken from our previous
observations (Shirley et al. 2003).

All observations except the C$^{34}$S 2$-$1 and 3$-$2 lines
were made at the 10.4 m telescope at the
Caltech Submillimeter Observatory
\footnote[1]{The CSO is operated by the California Institute of
Technology under funding from the National Science Foundation, contract
AST 90-15755.} (CSO). The C$^{34}$S 2$-$1 and 3$-$2 data were taken from
observations (Plume et al. 1997) at the IRAM 30-m telescope at Pico Veleta,
Spain.
The observing date, line frequency, beam size, main beam efficiency and
velocity resolution of each line are listed in Table 1.
The rest frequencies of
HCN $3-2$, H$^{13}$CN $3-2$, and C$^{34}$S lines have been updated 
(Ahrens et al. 2002, Gottlieb et al. 2003)
after our observations. We have corrected our data to the new
frequencies listed in Table 1. The position switching mode was used; 
offset positions were checked for HCN emission and found to be clear
($T_A^* < 0.5$ K). 
Pointing was checked periodically using planets and CO-bright stars.
The pointing accuracy was better than 6\arcsec .

\section{Results}

Table 2 lists our sources with their observed and derived parameters.
Self-absorption features appeared in many sources. We characterize
the line profiles by $v_{thick}$, the velocity at the peak of the
HCN $3-2$ line, $v_{thin}$, the velocity at the peak of the optically
thin tracer, and $\Delta v_{thin}$, the linewidth of the thin tracer.
Values were measured with a cursor or from a Gaussian fit, as appropriate.
Moving clockwise from the lower left, Figure 1 illustrates the sample's
range from a clearly blue profile (a) to a red profile (d).

A blue profile is a rather general feature of inflowing motion, but
it only arises for the right combination of critical density and opacity
of the line.
It is predicted for velocity fields that decrease with radius, as in
inside-out collapse models (Zhou \& Evans 1994), but also for more general
velocity fields (e.g., Myers et al. 1996).
In a simple model with two uniform layers approaching each other
(Myers et al. 1996), the shape of the profile is affected by both the
optical depth of the line and the inflow velocity of the front
layer (Myers et al. 2000).
In the most opaque sources, as shown in Figure 1(a), 
both HCN 3$-$2 and CS 5$-$4 lines show a blue
profile, but the HCN 3$-$2 line presents deeper absorption.
As opacity drops, only HCN shows an absorption
dip, as seen in Fig 1(b), and the less opaque CS
5$-$4 line becomes a single peak with a red shoulder. If the inflow
velocity is high enough, the red peak disappears, leaving only a flat wing
as seen in Fig 1(c).

The ratio of the blue peak to red peak
[$T^{*}_{R}(B)/T^{*}_{R}(R)$] is one measure of the
line asymmetry. A ``blue profile'' should have a stronger blue peak
than red peak: $T^{*}_{R}(B)/T^{*}_{R}(R) >1$ by a statistically significant
amount.
We calculated these ratios for the 18 sources with double-peaked spectra,
as listed in Table 2.
Eleven sources have a ratio of blue to red significantly
greater than unity, and 3 have a ratio significantly
less than unity; the other four cases do not differ from unity
by more than 1$\sigma$.
The overall trend is a sign of the blue predominance in the sample.

Not all line profiles from collapsing cores will show 2 peaks; some will
appear as lines skewed to the blue or lines with red shoulders.
An alternative definition (Mardones et al. 1997) is useful for these cases
as well.
A line can be identified as a blue profile if the peak of the optically
thick line is shifted blueward, with the velocity difference between
the peaks of the optically thick line and the optically thin line greater than
a quarter of the linewidth of the optically thin line:
$\delta v=(v_{thick}-v_{thin})/\Delta v_{thin} < -0.25$. A red profile
would have $\delta v > 0.25$.
The calculated $\delta v$ for our sample are listed in Table 2.
Twelve blue profiles and 6 red profiles were identified by this method, with
statistically significant ($\geq 1\sigma$) values of $\delta v$.
The distributions of $T^{*}_{R}(B)/T^{*}_{R}(R)$ and $\delta v$ are 
presented in Figure 2, which clearly shows the blue profile predominance.

The concept of the ``excess'' was introduced by Mardones et al. (1997) to
quantify the statistics of the line asymmetry in a survey:
$E=(N_{blue}-N_{red})/N_{total}$, where $N_{blue}$ and $N_{red}$ are
the numbers of blue and red profiles in the total sample of $N_{total}$
sources (28 in the present case). Using the line ratios or the
$\delta v$ measure, the excesses in our sample are $E=0.29$ and 0.21,
respectively.
For comparison, surveys of low mass star forming regions with HCO$^{+}$
$3-2$ found $E=0.30$, 0.31 and 0.31 for
Class $-$1, Class 0 and Class I samples, respectively
(Evans  2003).
Although a single line with a blue profile may be
explained by outflow or rotation, a large sample with a random distribution
of angles between the outflow or rotation axis and the line of sight
should result in a zero excess. A significant excess in a sample
is statistical evidence of inflow.
The probability that the observed number of blue and red sources  could result
from random sampling of a distribution that actually has equal numbers
in blue and red bins, with the total equal to the actual total, is
0.03 for the intensity criterion and 0.16 for the $\delta v$ criterion. 
The statistical significance is similar to that found for studies of
low-mass regions by Gregersen et al. (2000).

\section{Discussion}

Before accepting the inflow interpretation, we should examine other
possible explanations for the line profiles. One possibility is that
another molecular line contributes to the blue peak. The mean shift of the
peak for the 12 blue profiles is 
$\mean{v_{thick}-v_{thin}} = -2.18\pm 0.86$ \kms.  
Examination of
transition catalogs revealed only two lines within $\pm 5$ \kms\ of the
HCN line that have energies above ground corresponding to less than 300 K:
a line of glycolaldehyde shifted by $-0.4$ \kms; and a line of c-SiC$_3$
shifted by $+1.45$ \kms. Neither of these is likely to contribute
significantly to the observed line profile.
Accurate information on the hyperfine structure of the HCN $3-2$
transition has been reported recently (Ahrens et al. 2002).
The stronger components are all within $0.3$ \kms\ of the line center;
the two outer hyperfine components ($F = 2-2$ and $3-3$) lie at
$-2.35$ and $1.75$ \kms, respectively. These outer components could
affect the line profiles, but they have the same instrinsic line strengths,
at 0.037 of the total line strength. Ahrens et al. (2002) found the
blue-shifted component ($F = 2-2$) to be anomalously {\it weak} in
their observations of TMC-1. Thus, it seems unlikely that this component
is contributing strongly to the blue peak, but it remains a remote possibility.

For 13 sources where H$^{13}$CN is not available we used C$^{34}$S as the 
optically thin line. Based on the 9 sources for which
we have both H$^{13}$CN 3-2 
and C$^{34}$S 5-4 lines, the center velocity of H$^{13}$CN 3-2 has a mean shift 
of $-0.30$ \kms\ relative to the center of the C$^{34}$S 5-4 line, but the
linewidth of the H$^{13}$CN line is 1.24 times that of the C$^{34}$S line. 
These facts suggest that the H$^{13}$CN 3-2 is not optically thin, and
using it might make it harder to detect blue profiles. Indeed, using the
C$^{34}$S lines instead added two blue sources in the $\delta v$ statistic,
yielding an excess of 0.29, identical to that for the line ratio method.
We conclude that the choice of optically thin line does not have a major
effect on our statistics, but we have made a conservative choice by
using H$^{13}$CN lines where possible.

The self-absorption feature appears as long as the column density is high
enough (log N(CS)$\geq 14.3$ for this sample).
While HCN $3-2$ works well for such dense, opaque cores,
a different tracer may be needed for regions of lower column density.
The two-layer model predicts that $|\delta v|$ should increase
with optical depth (see Fig. 2 of Myers et al. 1996).
We observe that $|\delta v|$ increases with log N(CS), though the correlation
is weak.
The relation between $|\delta v|$ and column density supports the
idea that the blue profile only appears when the line has a suitable optical
depth and critical density. The excesses based on the $\delta v$ test for
the CS $5-4$ and $2-1$ lines in this sample are $E = -0.04$ and 0.05,
respectively. These negligible excesses are expected if the CS lines are less
opaque in the inflow region.

The current work is only a promising indication that inflow in regions forming
massive stars may be studied using HCN $3-2$ lines. The sources in this
sample span a wide range of distances; even if inflow motions are
present, they may be related to formation of individual stars in nearby
regions and formation of a cluster in more distant regions.
In future work, we will extend the sample to sources with smaller column
density. Maps will be made to test whether the blue profile peaks
toward the center or whether it is associated with outflows.
Sources that pass that test will become collapse candidates.
Using the models of density and temperature established by observations
of dust emission on the same sources (Mueller et al. 2002),
we could then test some models of collapse to learn whether we are
finally seeing gravitational infall in massive cores.

\acknowledgements
We are grateful to the staff of the CSO and to
K. Allers, C. Knez, and Y. Shirley for assistance with the observations.
We also thank Y. Shirley for use of his CS data before its publication.
We thank G. Fuller and P. Myers for helpful comments on an earlier draft,
and we thank the referee, M. Tafalla, for helpful comments and for alerting
us to the updated frequencies for C$^{34}$S.
This work was supported by NSF Grant AST-9988230 to the
University of Texas at Austin, and the state of Texas.

\begin{center}
\begin{table}
\caption{Observing Parameters }
\begin{tabular}{lccrrc}
\hline
Line & Date &$\nu$\tablenotemark{a}& $\theta_{mb}$ &$\eta_{mb}$ & $v_{res}$\\
      & (UT) & (GHz) & $('')$        &             &(\kms)\\\hline
HCN 3$-$2       & 2002 Jun & 265.886434 & 28.1 & 0.64 & 0.11 \\
HCN 3$-$2       & 2002 Dec & 265.886434 & 28.1 & 0.60 & 0.11 \\
H$^{13}$CN 3$-$2& 2002 Dec & 259.011814 & 28.9 & 0.60 & 0.11 \\
H$^{13}$CN 3$-$2& 2003 May & 259.011814 & 28.9 & 0.58 & 0.17 \\
C$^{34}$S 5$-$4&  2001 Jul & 241.016089 & 31.0 & 0.73 & 0.12 \\
C$^{34}$S 2$-$1&1991 Apr,Oct&96.4129495 & 25.0 & 0.60 & 0.31 \\
C$^{34}$S 3$-$2&1990 Jun   & 144.617101 & 17.0 & 0.60 & 0.21 \\
CS 5$-$4        & 1997 Apr & 244.935557 & 24.5 & 0.56 & 0.12 \\\hline
\end{tabular}\\
\tablenotetext{a}{The rest frequencies of HCN 3-2 and H$^{13}$CN 3$-$2
                   have been updated according to Ahrens et al. (2002).
  Those of CS and C$^{34}$S have been updated according to Gottlieb et al.
 (2003).}
\end{table}
\end{center}

\begin{center}
\begin{table}
\caption{Observed and derived parameters \tablenotemark{a}}
\scriptsize
\begin{tabular}{lccrrrcrllc}
\hline
Source & R.A. & Decl. & Dist & 
$v_{thick}$\tablenotemark{b}&$v_{thin}$\tablenotemark{c}
&$\Delta v_{thin}\tablenotemark{d}$
  & $\delta v$ & Pf\tablenotemark{e} & LogN\tablenotemark{f}
  &$T^{*}_{R}(B)/T^{*}_{R}(R)$\\
        &(1950.0)&(1950.0)& (kpc) &(\kms)&(\kms)&(\kms)&
        &   &(CS)&    \\\hline
RCW142      & 17 47 04.5 & $-$28 53 42 & 
2.0  &14.27(05)   &  17.24(05)$^{2}$  & 5.60(12)&$-$0.53(02) &B &15.2 
&4.18(69) \\
W28A2(1)    & 17 57 26.8 & $-$24 03 54 & 
2.6  &10.55(05)   &   9.17(08)$^{1}$  & 6.68(21)&   0.21(01) &N &15.7 &...\\
M8E         & 18 01 49.1 & $-$24 26 57 & 
1.8  &12.26(05)   &  10.65(10)$^{1}$  & 3.34(27)&   0.48(05) &R &14.8 
&0.95(08)\\
9.62+0.10   & 18 03 16.0 & $-$20 32 01 & 7.0  & 
7.11(06)   &   5.29(09)$^{1}$  & 5.89(26)&   0.31(02) &R &14.9 &0.26(04)\\
10.60$-$0.40& 18 07 30.7 & $-$19 56 28 & 6.5  &$-$6.51(05) 
&$-$4.51(08)$^{1}$  & 6.93(18)&$-$0.29(01) &B &15.6 &2.28(13)\\
19.61-0.23  & 18 24 50.1 & $-$11 58 22 & 
4.0  &40.52(05)   &  42.16(15)$^{3}$  & 7.18(29)&$-$0.23(02) &N &14.5 
&1.61(42)\\
23.95+0.16  & 18 31 40.8 & $-$07 57 17 & 
5.8  &78.75(17)   &  80.15(09)$^{2}$  & 2.39(22)&$-$0.59(11) &B 
& ... &1.17(25)\\
W43S        & 18 43 26.7 & $-$02 42 40 & 
8.5  &96.83(05)   &  98.17(08)$^{1}$  & 6.00(23)&$-$0.22(02) & N 
&14.0 &1.68(15)\\
W44         & 18 50 46.1 &   +01 11 11 & 3.7  &54.46(05)   
&  57.20(08)$^{1}$  & 5.78(10)&$-$0.47(01) &B &14.3 &3.31(31)\\
35.58-0.03  & 18 53 51.4 &   +02 16 29 &10.2  &53.79(06)   
&  54.04(15)$^{3}$  & 5.16(27)&$-$0.05(03) &N &14.3 &...\\
48.61+0.02  & 19 18 13.1 &   +13 49 44 &11.8  &17.87(05)   
&  18.03(11)$^{2}$  & 2.34(21)&$-$0.07(05) &N &14.1 &...\\
59.78+0.06  & 19 41 04.2 &   +23 36 42 & 2.2  &21.89(05)   
&  22.63(07)$^{2}$  & 2.21(23)&$-$0.33(05) &B &14.1 &...\\
S88B        & 19 44 42.0 &   +25 05 30 & 2.1  &22.30(11)   
&  21.42(05)$^{2}$  & 2.36(13)&   0.37(05) &R &13.8 &...\\
ON2S        & 20 19 48.9 &   +37 15 52 & 5.5  &$-$0.05(05) 
&$-$0.90(06)$^{2}$  & 3.65(14)&   0.23(02) &N &14.3*&...\\
W75N        & 20 36 50.5 &   +42 27 01 & 3.0  &11.70(10)   
&   9.26(08)$^{1}$  & 5.47(07)&   0.45(02) &R &14.4*&0.71(04)\\  
DR21S       & 20 37 13.8 &   +42 08 52 & 3.0  &$-$5.68(05) 
&$-$2.18(10)$^{4}$  & 4.38(13)&$-$0.80(03) &B &14.4*&1.99(11)\\
W75(OH)     & 20 37 14.0 &   +42 12 12 & 3.0  &$-$6.01(05) 
&$-$3.32(08)$^{1}$  & 5.95(17)&$-$0.45(04) &B &14.8*&1.29(06)\\
CEPA        & 22 54 19.2 &   +61 45 44 & 0.7  &$-$12.42(13)
&$-$10.39(09)$^{2}$ & 3.68(19)&$-$0.55(05) &B &14.5 &1.06(07)\\
S158        & 23 11 36.1 &   +61 10 30 & 2.8  &$-$59.46(11)
&$-$56.23(12)$^{1}$ & 6.15(29)&$-$0.53(03) &B &14.6*&1.78(20)\\
S157        & 23 13 53.1 &   +59 45 18 & 2.5  &$-$45.05(11)
&$-$43.82(10)$^{4}$ & 2.56(20)&$-$0.48(07) &B &13.9*&1.39(33)\\
121.30+0.66   & 00 33 53.3 &   +63 12 32 & 1.2  &$-$18.55(05)
&$-$17.77(09)$^{1}$ & 3.09(23)&$-$0.25(04) &N &14.0*&1.33(14)\\
123.07$-$6.31 & 00 49 29.2 &   +56 17 36 & 2.2  &$-$32.67(05)
&$-$30.84(11)$^{1}$ & 4.68(28)&$-$0.39(04) &B &14.0*&1.27(28)\\
W3(OH)      & 02 23 17.3 &   +61 38 58 & 2.4  &$-$49.39(05)
&$-$46.85(08)$^{1}$ & 5.18(20)&$-$0.49(03) &B &15.0*&1.44(15)\\
136.38+2.27 & 02 46 11.7 &   +61 47 34 & 4.5  &$-$42.18(18)
&$-$42.35(10)$^{4}$ & 2.03(16)&   0.08(10) &N &13.6*&...\\
S231        & 05 35 51.3 &   +35 44 16 & 2.3  &$-$14.22(05)
&$-$16.41(05)$^{1}$ & 3.72(14)&   0.59(03) &R &14.3 &0.34(07)\\
S235        & 05 37 31.8 &   +35 40 18 & 1.6  &$-$17.26(05)
&$-$17.44(04)$^{1}$ & 2.73(10)&   0.07(02) &N &14.3*&...\\
S252A       & 06 05 36.5 &   +20 39 34 & 1.5  &    9.90(05)
&    9.10(06)$^{1}$ & 2.57(14)&   0.31(04) &R & ... &....\\
S255$/$7    & 06 09 58.2 &   +18 00 17 & 1.3  &    6.80(10)
&    6.98(10)$^{4}$ & 2.57(10)&  -0.07(06) &N &14.3*&...\\\hline

\end{tabular}\\
\tablenotetext{a}{Units of right ascension are hours, minutes, and seconds, and
units for declination are degrees, arcminutes, and arcseconds.
Distances are taken from Shirley et al. (2003) and Mueller er al. (2002).
Quantities in parentheses give the uncertainties in units of 0.01.}
\tablenotetext{b}{The velocity of the peak of the optically
thick line, in this case HCN 3$-$2.}
\tablenotetext{c}{The velocity of the peak of the optically thin
line, whether H$^{13}$CN 3$-$2 or C$^{34}$S 5$-$4, 2$-$1, and 3$-$2, indicated
by 1, 2, 3 and 4 respectively.}
\tablenotetext{d}{FWHM of the optically thin line. For 3 sources (W28A2(1),
9.62+0.10 and W44), we had to mask a strong wing component to get an
accurate line width measurement.}
\tablenotetext{e}{Profile of the line judged from $\delta v$. B denotes
blue profile and R denotes red profile. N denotes neither blue or red profile.}
\tablenotetext{f}{Taken from Plume et al. 1997, N is the column density
derived from an LVG model fitted to several CS transitions.
Densities with * are calculated from fitting several C$^{34}$S transitions,
corrected by the average difference of the log of the column densities,
$\mean{{\rm log}N({\rm CS})-{\rm log}N({\rm C}^{34}{\rm S})}=0.83\pm 0.27 $.}

\end{table}
\end{center}

\begin{figure}[hbt!]
\epsscale{0.8}
\plotone{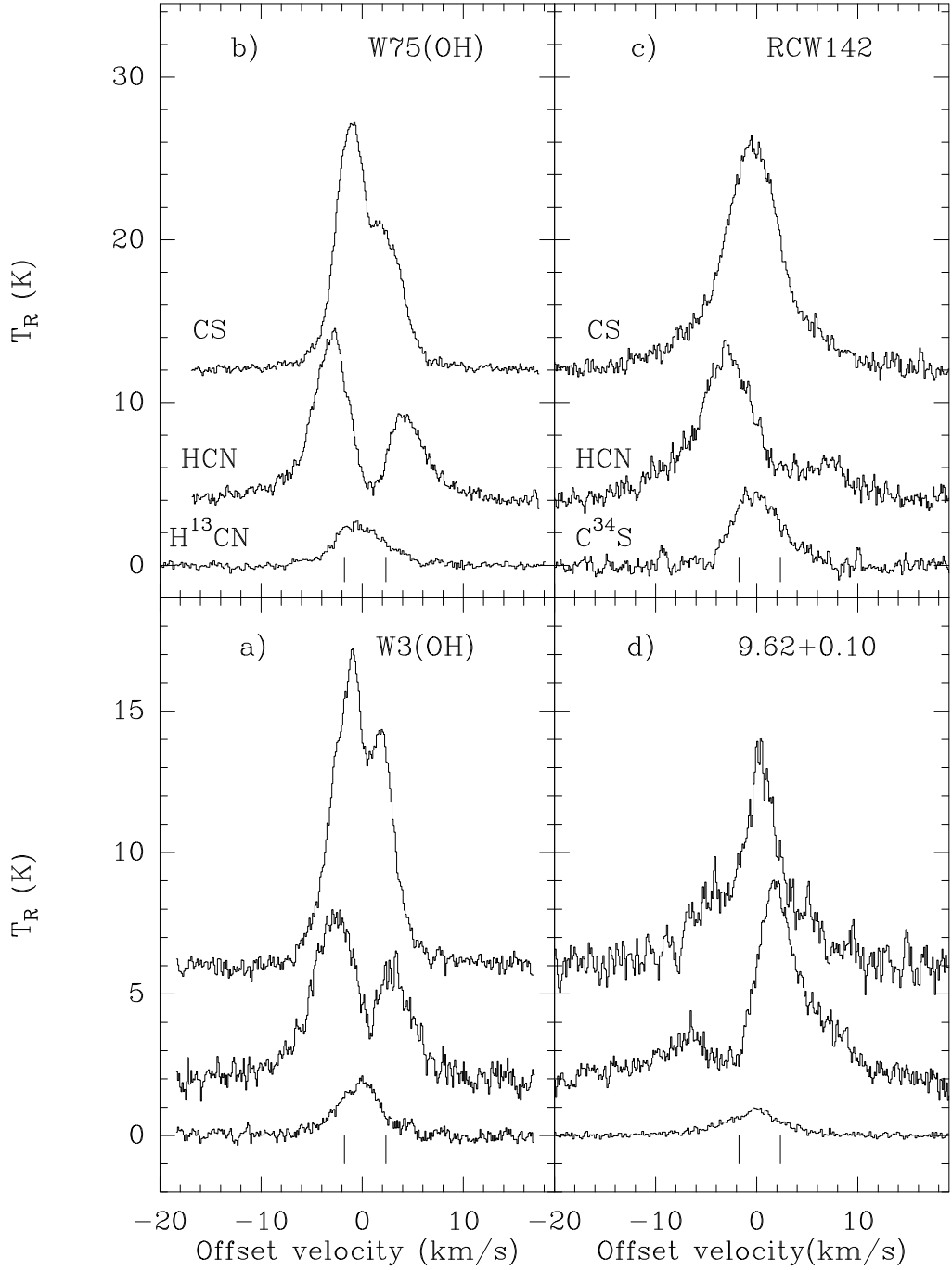}
\caption{Line profiles of HCN $3-2$ (middle lines), CS $5-4$ (upper lines)
  and optically thin lines (lower lines,
  H$^{13}$CN $3-2$ for a), b), d) and C$^{34}$S $5-4$ for c). HCN $3-2$ and
  CS $5-4$ lines have been displaced upward for clarity.
  All lines are plotted with the velocity relative to the
optically thin line's central velocity. The two vertical lines below the
spectra indicate the location of the outermost hyperfine components of the
HCN $3-2$ line.
The radiation temperature $T_{R}$ has been corrected for main beam efficiency.}
\label{f1}
\end{figure}

\begin{figure}[hbt!]
\plotone{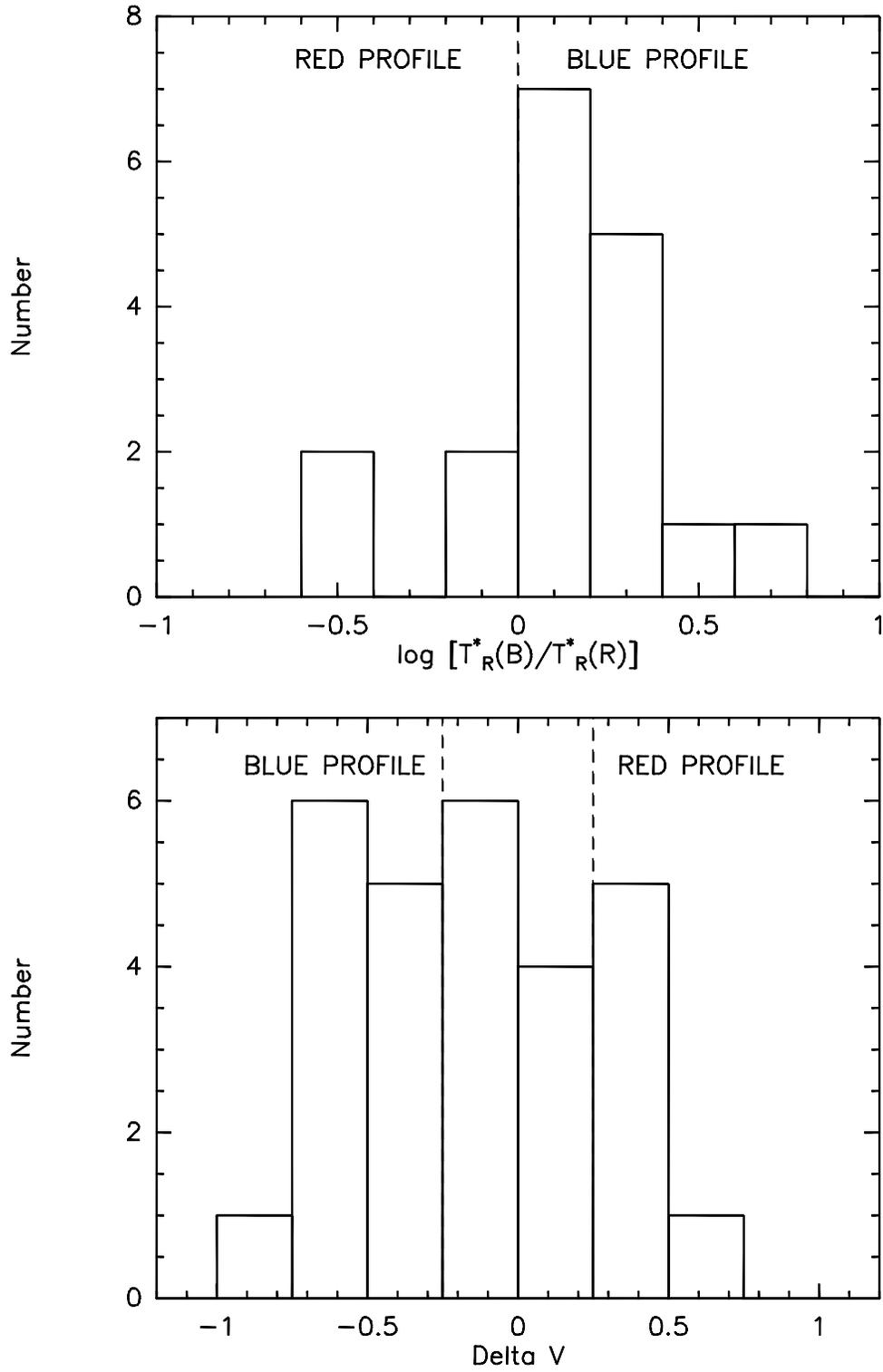}
\caption{Distribution of log[$T^{*}_{R}(B)/T^{*}_{R}(R)$] for the 18 
 double-peaked sources (top plot) and distribution of $\delta v$ 
 for all the 28 sources in the sample (bottom plot). Note that only 11
of the 14 blue and 3 of the 4 red sources in the top plot have ratios
greater or less than unity by more than 1 $\sigma$.}
\label{f2}
\end{figure}

\end{document}